\documentclass[11pt,a4paper]{article}
\pdfoutput=1
\usepackage{jheppub}
\usepackage{amsfonts}
\usepackage{bbm}
\usepackage{graphicx}
\usepackage{epsfig}
\usepackage{caption}
\usepackage{subcaption}
\usepackage{bigints}
\usepackage[normalem]{ulem}
\usepackage{comment}
\usepackage{mciteplus}
\usepackage{epsfig}

\newcommand{\nn}{\nonumber}

\def\ben{\begin{eqnarray}}
\def\een{\end{eqnarray}}
\def\be{\begin{equation}}
\def\ee{\end{equation}}
\def\ba{\begin{align}}
\def\ea{\end{align}}

\newcommand{\bem}{\begin{pmatrix}}
\newcommand{\eem}{\end{pmatrix}}

\def\={\;  = \;}
\def\+{\, + \,}

\def\bar{\overline}

\def\rt2{\sqrt{2}}

\title{Stability, Tunneling and Flux Changing de Sitter Transitions in the Large Volume String Scenario}

\author{S. de Alwis${}^{1,2}$, R. Gupta${}^2$, E. Hatefi${}^2 $ and F.~Quevedo${}^{2,3}$}

\affiliation{$^1$  UCB 390 Physics Dept. University of Colorado, Boulder CO 80309, USA\\
$^2$ Abdus Salam ICTP, Strada Costiera 11, Trieste 34014, Italy \\
$^3$ DAMTP, University of Cambridge, Wilberforce Road,  Cambridge, CB3 0WA, UK}

\emailAdd{dealwiss@gmail.com, rgupta@ictp.it, ehatefi@ictp.it, F.Quevedo@damtp.cam.ac.uk}

\abstract{We study the non-perturbative stability of the Large Volume Scenario (LVS) of IIB string compactifications, by analysing transitions mediated by the Brown-Teitelboim (BT) brane nucleations and by  Coleman De Luccia tunneling (CDL). We find that,  as long as the effective field theory description holds, the LVS AdS minima are stable despite being non-supersymmetric.  This opens the possibility of having a CFT dual. Metastable de Sitter vacua behave differently depending on the uplifting mechanism. We find explicit expressions for the different decay rates in terms of exponentials of the volume.
Among the transitions of dS to dS those with increasing volume and decreasing vacuum energy are preferred, though dS decays to AdS (big-crunch sinks) have higher probability. 
However, the probability of transitions via the CDL mechanism to decompactification are exponentially much larger compared to these. 
The  BT decays correspond to flux/D3 brane transitions mediated by the nucleation of D5/NS5 branes. We compare our results with previous analysis for KKLT, type IIA, and 6D Einstein-Maxwell studies. In particular we find  no indication for a bubble of nothing decay.}

\begin{document}

\vspace*{-2cm} 
\begin{flushright}
{\tt  DAMTP-2013-42 } 
\end{flushright}

\maketitle
\section{Introduction}

Successful moduli stabilisation of string flux compactifications generically lead to a large number of 4-dimensional scalar potentials, with local minima corresponding to Anti-de Sitter (AdS), de Sitter (dS), or Minkowski space-time in 4-dimensions. In the case of supersymmetric minima these vacua tend to be also non-perturbatively stable. But for non-supersymmetric minima non-perturbative instabilities naturally arise. These may be due to tunneling \cite{cdl}, brane nucleation \cite{bt}\ and other transitions such as to bubbles of nothing \cite{witten}. For KKLT  moduli stabilisation \cite{kklt} before uplifting, the minimum is supersymmetric and stability is essentially guaranteed. Once an uplifting mechanism is added to AdS space, giving dS space, supersymmetry is broken and the minimum becomes metastable, decaying by tunneling to the decompactified 10-dimensional Minkowski space, but also by  brane nucleation to other vacua with different values of the vacuum energy  obtained from varying values of the fluxes and then, as long as effective field theory description is valid, populating the landscape.

In this article we consider the transitions for the large volume scenario (LVS)  \cite{lvs}\ in which K\"ahler moduli are stabilised yielding  an exponentially large volume. Contrary to KKLT the AdS minima are already non-supersymmetric and therefore subject to potential instabilities.  In particular there is a potential decay to a bubble of nothing.  But also to other minima with different values of the fluxes such as the KKLT minimum when it exists. Uplifting to dS has been done in several ways and we consider the decay products in each case.  We  consider both decay by the Brown-Teitelboim mechanism corresponding to brane nucleation and  by tunneling following the Coleman-de Luccia formalism \footnote{Some other potential instabilities in the presence of dense matter have been studied \cite{Conlon}. In this paper we are considering decay via real valued instantons however complex valued instantons have also been studied \cite{Hartle}. }. We summarise our results as follows:

\begin{itemize}

\item{} We find that the (non-supersymmetric) AdS minimum in the LVS is stable as long as the effective field theory is valid. This in particular indicates that despite the vacuum not being supersymmetric, there is no obstacle for having a CFT dual and then a proper non-perturbative description.

\item{} The dS uplifted vacuum is metastable, tunneling as usual to decompactified 10-dimensional minimum but also decaying by bubble nucleation to vacua with reducing number of fluxes. The transitions are due to the nucleation  of D5/NS5 branes similar  to but different in detail to the case studied in \cite{kpv}.

\item{} All decay rates go like $e^{-{\cal V}^3}$ to leading order in the volume ${\cal V}$ expansion. However, the decay rate of dS to dS is much suppressed than the decompactification decay via CDL and is also suppressed compared to the decay to AdS (big-crunch sinks) \cite{sinkskklt,maldacena}.  All decay rates are much larger than the Poincare recurrence rate.

\item{} The dS to dS decays depend on the up-lifting terms. We consider the two general classes of uplifts that have been proposed in the literature, based on whether or not the uplift depends directly on the flux superpotential. We find in both cases  higher probability transitions towards increasing the volume and decreasing the cosmological constant, although in one case with increasing and the other one with deceasing superpotential.

\item{} Within effective field theory we find no indication of the bubble of nothing decay \cite{witten, Yang, blanco, browndahlen} if this follows the mechanism proposed in \cite{browndahlen}.

\end{itemize}

It is worth emphasising that our knowledge of the string landscape is very limited since it relies on the validity  and structure of an effective field theory (EFT) in 4 dimensions \footnote{See \cite{banks,shanta} for  criticisms of the use of EFT treatment and \cite{polchinski} for an answer to some of the critics.}. For a given value of the integer fluxes it usually gives rise to one or a few local (A)dS minima  as well as the overall minimum corresponding to decompactified 10 dimensions. The decay from a dS minimum to the decompactified one can be done by standard CDL (or Hawking Moss (HM)) transition within one EFT. However the great degeneracy of minima is generated by varying the quantised value of the fluxes which give rise to  another 4D scalar potential with its own collection of local minima. Transitions between vacua with different values of the fluxes cannot be described by one single EFT. Therefore the standard CDL and HM mechanisms cannot be used. For these flux transitions the proper approach is through the Brown-Teitelboim (BT) process corresponding to a bubble nucleated flux/brane transitions. Even though in practice this usually provides similar results as the CDL mechanism in the thin wall approximation \cite{frey}, we prefer to perform the analysis directly using BT. We devote the next section to a review of the latter.

Section 3 is dedicated to a brief review of the relevant aspects of flux compactifications in IIB string theory, presenting the relevant quantities needed for our analysis, such as the flux superpotential and tension of the branes. In particular  the BT mechanism corresponds to flux/D3-brane transitions caused by the nucleation of D5/NS5 branes. 
In the last part of this section  we briefly review the large volume scenario (LVS) emphasising the similarities and differences with the KKLT scenario (exponentially large volume, non-supersymmetric AdS minimum and generic ${\cal{O}} (1)$ flux superpotential.). 

Section 4 describes the stability analysis for  unlifted AdS minima and  the two general classes of uplift terms that have been considered in the literature. In the first class, the uplift term is proportional to $|W_0|^2$  as in the D-term induced F-term uplift. The second class has an uplift term independent of the flux superpotential (as occurs in the original anti D3 brane uplift and in non-perturbative superpotential from branes at singularities). 

We work with general expressions for these uplift terms in order to capture general classes of mechanisms rather than committing to one single proposal, since this is the most model-dependent component of the moduli stabilisation process. We then discuss  how the BT mechanism applies to the dS/AdS transitions for the LVS, establishing the main results of this article mentioned above. We end in section 6 with a general comparison between our scenario and others that have been discussed in the literature, such as KKLT, type IIA and 6-dimensional Einstein-Maxwell systems, in which also a landscape  of flux induced vacua exists, but with different physical properties from the LVS.

\section{\bf Vacuum decay rates : General discussion}

\subsection{The Brown-Teitelboim Mechanism}

The Brown-Teitelboim (BT) mechanism \cite{bt}\ describes changing the cosmological constant (CC) by a dynamical process of brane nucleation. In this mechanism a field initially in a metastable state,  with some vacuum energy, undergoes a transition to another vacuum state with different vacuum energy. In the initial state there is a spontaneous nucleation of a domain wall which expands and divides the space-time into two regions having a different value of fluxes and CC. \\
The probability per unit volume per unit time for brane nucleation in a  vacuum with CC $\Lambda_o$ for  decay to a  vacuum with CC  $\Lambda_i$ is given by
\be\label{probP}
P\sim e^{-B},\quad B=S[instanton]-S[background],
\ee
where $S$ is the Euclidean action. 

In field theory there is a similar process, described by Colemann and De Luccia (CDL) \cite{cdl}, of decay of false vacuum to true vacuum.  However there is a very important difference between CDL and BT processes. The former is a field theory process which describes tunneling between two minima of a potential and stops once the field reaches in its true minimum. However the membrane nucleation will always be (may be) repeated for dS (AdS) with the inside value of flux and CC  now become a background configuration. In this sense the BT process is more suitable for describing the string landscape.\\

The probability per unit volume per unit time for brane nucleation is given in terms of $B$.  In \cite{bt} one has a universal expression for B valid for any decay. The corresponding $B$ is given by
\be\label{probB}
B=2\pi^2\bar\rho^3T+12\pi^2\left\{\frac{1}{\Lambda_i}\left[\sigma_i\left(1-\frac{\Lambda_i}{3}\bar\rho^2\right)^{3/2}-1\right]-\frac{1}{\Lambda_o}\left[\sigma_o\left(1-\frac{\Lambda_o}{3}\bar\rho^2\right)^{3/2}-1\right]\right\}.
\ee
Here $\sigma_{o/i}=\pm 1$  is determined from
\be\label{signsigma}
\sigma_o=Sign\left[\frac{\epsilon}{3}-\frac{T^2}{4}\right],\quad\sigma_i=Sign\left[\frac{\epsilon}{3}+\frac{T^2}{4}\right],
\ee
 $T$ is the tension of the bubble wall and $\epsilon$ is defined as
\be
\epsilon=\Lambda_o-\Lambda_i.
\ee
It is also obvious from (\ref{signsigma}) that
\be
\sigma_i\geq\sigma_o.
\ee
The choice of $\sigma_{o/i}$ gives many possibilities of decay.  As we will see later, the choices which are relevant to us are
\be
\sigma_o=\pm 1,\quad \sigma_i=+1.
\ee 
Here $\bar\rho$ is the size of the bubble and is determined by extremizing B,
\be\label{barrho}
\bar\rho=\left\{\frac{\Lambda_o}{3}+\frac{1}{T^2}\left[\frac{\epsilon}{3}-\frac{T^2}{4}\right]^2\right\}^{-1/2}.
\ee
From (\ref{barrho}), we get the following condition
\be\label{nucl.condition}
\left[\frac{\epsilon}{3}-\frac{T^2}{4}\right]^2\geq -\frac{T^2\Lambda_o}{3}.
\ee
Thus if we start with de Sitter space for which $\Lambda_o>0$, then this condition is automatically satisfied. However for $\Lambda_o<0$ which is the case of AdS space, this inequality has to be satisfied in order to have a brane nucleation.

The outcomes of the BT brane nucleation process are:
\begin{enumerate} 
\item{} As long as the the initial space-time is de Sitter, there will always be a nucleation of a brane. Brane nucleation can increase or decrease the cosmological constant (CC). However it is very simple to see that the brane nucleation which decreases the CC occurs with greater probability. In fact in the limit when the tension of the brane is very small, the brane nucleation which increases the CC is highly suppressed.
 \item{} In Anti-de Sitter space as long as (\ref{nucl.condition}) is satisfied, the brane nucleation always occurs and it decreases the CC. 

\end{enumerate}

\subsection {Different classes of BT transitions} 

According to Brown-Teitelboim, there are 5  possible decays from de-Sitter and one from Anti de Sitter.\\ \\
{\bf Case 1:}${\bf \quad\sigma_i=+1, \quad\Lambda_{o(i)}>0 \quad dS\rightarrow dS}$\\
In this case we have
\be
B=2\pi^2\bar\rho^3T+12\pi^2\left\{\frac{1}{\Lambda_i}\left[\left(1-\frac{\Lambda_i}{3}\bar\rho^2\right)^{3/2}-1\right]-\frac{1}{\Lambda_o}\left[\sigma_o\left(1-\frac{\Lambda_o}{3}\bar\rho^2\right)^{3/2}-1\right]\right\}\ .
\ee

We have two possiblities here, \\
{\bf (i)}  ${\bf\sigma_o=1,}$\\
In this case we have
\be
B=2\pi^2\bar\rho^3T+12\pi^2\left\{\frac{1}{\Lambda_i}\left[\left(1-\frac{\Lambda_i}{3}\bar\rho^2\right)^{3/2}-1\right]-\frac{1}{\Lambda_o}\left[\left(1-\frac{\Lambda_o}{3}\bar\rho^2\right)^{3/2}-1\right]\right\}.
\ee
Since $\sigma_{o/i}=+1$, (\ref{signsigma}) implies that $\epsilon>0$. This will describe a process by which bubble nucleation reduces the cosmological constant. However the tension of the brane has an upper bound coming from the condition $\sigma_o=+1$ which is
\be
T^2<\frac{4\epsilon}{3}.
\ee
Thus when the tension satisfies the above bound, the bubble nucleation will reduce the cosmological constant.\\

{\bf (ii)}  ${\bf \sigma_o=-1,}$\\
In this case we have
\be\label{Bplusminus}
B=2\pi^2\bar\rho^3T+12\pi^2\left\{\frac{1}{\Lambda_i}\left[\left(1-\frac{\Lambda_i}{3}\bar\rho^2\right)^{3/2}-1\right]+\frac{1}{\Lambda_o}\left[\left(1-\frac{\Lambda_o}{3}\bar\rho^2\right)^{3/2}+1\right]\right\}.
\ee
Since $\sigma_o=-1$ and $\sigma_i=1$, there are two possibilities depending on the sign of $\epsilon$. In both cases, the tension has lower bound $T^2>\frac{4|\epsilon|}{3}.$

In the case when $\epsilon>0$, the bubble nucleation will reduce the cosmological constant. In the case when $\epsilon<0$, the bubble nucleation will increase the cosmological constant.\\
{\bf Case 2:} \quad ${\bf \sigma_i=-1,\quad \Lambda_{o(i)}>0,\quad dS\rightarrow dS}$.\\
In this case we have only one possibility with $\sigma_o=-1$. \\
\be
B=2\pi^2\bar\rho^3T-12\pi^2\left\{\frac{1}{\Lambda_i}\left[\left(1-\frac{\Lambda_i}{3}\bar\rho^2\right)^{3/2}+1\right]-\frac{1}{\Lambda_o}\left[\left(1-\frac{\Lambda_o}{3}\bar\rho^2\right)^{3/2}+1\right]\right\}.
\ee

Since in this case $\epsilon<0$, the bubble nucleation will increase the cosmological constant. In this case the  tension has lower bound $\frac{T^2}{4}<\frac{|\epsilon|}{3}$.\\
{\bf Case 3:}\quad
${\bf \sigma_i=+1,\Lambda_o>0,\Lambda_i<0, \quad dS\rightarrow AdS/flat}$\\
{\bf (i)} \quad ${\bf \sigma_o=1.}$\\
In this case we have
\be
B=2\pi^2\bar\rho^3T-12\pi^2\left\{\frac{1}{|\Lambda_i|}\left[\left(1+\frac{|\Lambda_i|}{3}\bar\rho^2\right)^{3/2}-1\right]+\frac{1}{\Lambda_o}\left[\left(1-\frac{\Lambda_o}{3}\bar\rho^2\right)^{3/2}-1\right]\right\}.
\ee
In this case bubble nucleation will decrease the cosmological constant. The tension has lower bound $T^2<\frac{4\epsilon}{3}.$\\
When $\Lambda_i\rightarrow 0$, we get
\be
B=\frac{32\pi^2}{T^2\left(1+\frac{T_c^2}{T^2}\right)^3}\left[\left(2-\frac{T_c^2}{T^2}\right)^2+\frac{T_c^2}{T^2}\right],\quad T_c^2=\frac{4\Lambda_o}{3}
\ee

{\bf (ii)} \quad ${\bf \sigma_o=-1.}$\\
\be
B=2\pi^2\bar\rho^3T+12\pi^2\left\{-\frac{1}{|\Lambda_i|}\left[\left(1+\frac{|\Lambda_i|}{3}\bar\rho^2\right)^{3/2}-1\right]+\frac{1}{\Lambda_o}\left[\left(1-\frac{\Lambda_o}{3}\bar\rho^2\right)^{3/2}+1\right]\right\}
\ee
In this case bubble nucleation decreases the cosmological constant. The tension has lower bound $T^2>\frac{4\epsilon}{3}$.\\
We consider decay from dS to flat space $\Lambda_i\rightarrow 0$, we get
\ben
\bar\rho &=&\frac{4}{T\left(1+\frac{T_c^2}{T^2}\right)},\qquad \epsilon=\frac{3T_c^2}{4}\nonumber \\
B &=&\frac{32\pi^2}{T_c^2 \left(1+\frac{T_c^2}{T^2}\right)^2}
\een
{\bf Case 4:}\\
${\bf \sigma_o=\sigma_i=+1,\quad \Lambda_{o(i)}<0,\quad AdS\rightarrow AdS}$\\
In this case we have
\be
B=2\pi^2\bar\rho^3T+12\pi^2\left\{-\frac{1}{|\Lambda_i|}\left[\left(1+\frac{|\Lambda_i|}{3}\bar\rho^2\right)^{3/2}-1\right]+\frac{1}{|\Lambda_o|}\left[\left(1+\frac{|\Lambda_o|}{3}\bar\rho^2\right)^{3/2}-1\right]\right\}
\ee
In \cite{bt} it has been shown that this is the only case when we have decay between two AdS spaces. From (\ref{signsigma}) we see that in this case $\frac{\epsilon}{3}>\frac{T^2}{4}$. Also we have inequality (\ref{nucl.condition}) to satisfy in order to have decay. This inequality simplifies to 
\ben
&&T^2+4\sqrt{\frac{|\Lambda_o|}{3}}T-\frac{4\epsilon}{3}\leq 0\nonumber \\
&&T\leq \sqrt{\frac{4}{3}}(\sqrt{(|\Lambda_o|+\epsilon)}-\sqrt{{|\Lambda_o|}})
\een

For decay from flat space $(\Lambda_o=0)$ to AdS space, we get
\be
B=2\pi^2\bar\rho^3T+6\pi^2\bar\rho^2-\frac{12\pi^2}{|\Lambda_i|}\left[\left(1+\frac{|\Lambda_i|}{3}\bar\rho^2\right)^{3/2}-1\right]
\ee
Putting
\be
\bar\rho=\frac{T}{|\frac{T^2}{4}-\frac{\epsilon}{3}|}
\ee
We get (note $\epsilon =-|\Lambda_i|$ here)
\be
B=\frac{27\pi^2T^4}{2\epsilon^3}\frac{1}{[1-3T^2/4\epsilon]^2}.
\ee

\section{Transitions and Brane Nucleation in String Theory}

\subsection{Flux compactifications and Flux/Brane Transitions}

The minima that we consider are in large volume regions of
moduli space so that we can effectively ignore warping. Following \cite{gkp} we consider the following
metric ansatz:
\begin{equation}
ds^{2}=e^{\phi/2}[e^{-6u(x)}g_{\mu\nu}(x)dx^{\mu}dx^{\nu}+e^{2u(x)}g_{mn}dy^{m}dy^{n}],\label{eq:metric}
\end{equation}
where $ds^{2}$ is the string frame metric, $\phi$ is the (10 D)
dilaton, $g_{\mu\nu}$ is the 4D Einstein frame metric, and $g_{mn}$
is a fiducial metric on the Calabi-Yau Orientifold (CYO) $X$. Here
$e^{u}$ is the radius of the internal space and ${\cal V}\sim e^{6u}$
is the volume of the internal space.  The string
scale (tension) in string frame is $M_{s}^{2}=1/2\pi\alpha'$ and when the moduli are stabilised (at a minimum denoted by $|_0$),
\begin{equation}
\frac{M_{s}^{2}}{M_{P}^{2}}=\frac{1}{2}e^{\phi/2}e^{-6u}|_{0}.\label{eq:stringscale}
\end{equation}

As discussed in \cite{gkp} the superpotential can be expanded in terms of  integers that characterize the NSNS (RR) fluxes $n_i$($m_i$) threading the 3 cycles ($A$ and $B$) of the CYO. We introduce  homogeneous coordinates on the complex structure moduli in the usual way as integrals of the holomorphic three form $\Omega$ over these cycles and choose  $z^0=1$ (fixing the scale of  $\Omega$). This gives
\begin{equation}
W_{flux}=\int_{X}G_{3}\wedge\Omega=\sum_{i=0}^{h_{12}}[(n^i_{A}-iSm_{A}^{i}){\cal G}_{i}(z)-(n_{i}^{B}-iSm_{i}^{B})z^{i}]\equiv A(n,,z)+B(m,z)S.\label{eq:Wflux}
\end{equation}
Here $S$ is the axio-dilaton field with $Re S =e^{-\phi}$ in type IIB case\footnote{ For more details in the type IIB case see section 3.2.}.\\
The effect of elementary flux changing transitions with all fluxes except for the
one indicated unchanged are then given as follows \footnote{The discussion in this subsection is based on \cite{sda}. For an earlier work discussing the role of the same configuration in relating vacua of different numbers of supersymmetry see \cite{Kachru:2002ns}.} :
\begin{eqnarray*}
n_{i}^{B}\rightarrow n_{i}^{B}\pm1 &  & \Delta W=\mp z^{i},\\
m_{i}^{B}\rightarrow m_{i}^{B}\pm1 &  & \Delta W=\pm iSz^{i},\\
n_{A}^{i}\rightarrow n_{A}^{i}\pm1 &  & \Delta W=\pm{\cal G}_{i},\\
n_{A}^{i}\rightarrow n_{A}^{i}\pm1 &  & \Delta W=\mp iS{\cal G}_{i}.
\end{eqnarray*}
Any elementary transition in the geometric weak coupling regime will
change $W$ by $O(1)$. Following the logic of LVS compactifications \cite{lvs}\footnote{See next section for a short review.}, to leading order in the inverse volume expansion  we may neglect the  non-perturbative (NP) term in solving
$D_{S}W=0,\, D_{i}W=0,\, i=1,\ldots,h_{12}$. Solving the first gives
\begin{equation}
\bar{S}=\bar{S}_{0}\equiv\frac{A(n,z)}{B(m,z)}\label{eq:Sbar}
\end{equation}
Plugging into the second set of equations, gives
\begin{equation}
\sum_{i=1}^{h_{12}}\left[\left(n_{A}^{i}{\cal G}_{ij}(z)-n_{j}^{B}\right)+i\frac{\bar{A}(n,\bar{z})}{\bar{B}(m,\bar{z})}\left(m_j^B-m_{A}^{i}{\cal G}_{ij}(z)\right)\right]+K_{j}W=0\label{eq:zeqns}
\end{equation}
These equations  (and their complex conjugates) are a set of $2h_{12}+2$ equations for $2h_{12}+2$
real variables ${\rm Re} z, {\rm Im} z, {\rm Re}  S, {\rm Im}  S$. A solution is not guaranteed for arbitrary
sets of flux integers - we need to scan over integer sets to get $z=z_{0}(n,m)$
and then $S=S(n,m)$. In other words, only some sets of elementary transitions
will lead to potentials with the $z$'s (and $S$) stabilized supersymmetrically in the region consistent with an effective field theory analysis. 

Note that in type IIB string theory the total number of fluxes is $4h_{21}+4$. So after fixing the $z^{i}$'s
and $S$ there are $2h_{21}+2$ fluxes left that can generate a discretuum that can be used to 
find a small  CC.
Hence there are  transitions in this theory that  just change the CC without having any effect on the moduli or the dilaton.

In order to change the flux through the three cycles and create a new 4D vacuum with a different CC we need to nucleate a five-brane. This is the only type of BPS brane in IIB string compactifications that can accomplish this since it needs to form an $S^2$ in the non-compact 3D space - thus dividing it into two domains with different physics - and wrap a three cycle in the Calabi-Yau orientifold. The brane can be either a NSNS or a D-brane - as long as the transitions we are discussing do not change the string coupling drastically from $g_s\lesssim 1$, the probability of nucleating either should be of similar magnitude. For concreteness we will discuss the D5 brane case.

In the string frame (in  units where $2\pi\sqrt{\alpha'}=1$.) the tension of a D5 brane is given by 
\be
T_5^{s}=2\pi e^{-\phi}
\ee
So the  action of the D5 brane in the probe limit is
\be
S_{D5}=2\pi\int d^3xd^3ye^{-\phi}\sqrt{G}
\ee
Here $G$ is the string frame metric related to Einstein metric by $G_{MN}=e^{\phi/2}g_{MN}$ and $\phi$ is the dilaton.\\
The D5 action in Einstein frame is
\ben
S_{D5}&=&2\pi\int d^3xd^3ye^{\phi/2}\sqrt{g}e^{-6u}\nonumber \\
&=& 2\pi\int_\Sigma d^3y\sqrt{g'}\int d^3xe^{\phi/2}\sqrt{g^{(3)}}e^{-6u}.
\een
Here $\Sigma$ is a 3-cycle which the brane is wrapping and $g'_{mn}$ is the induced metric along three cycle.\\
Thus the effective tension of the bubble wall is given by
\be
T_{wall}=2\pi\int_\Sigma d^3y\sqrt{g'}e^{\phi/2}e^{-6u}
\ee
The D5-brane tension, which wraps $r_{A}^{i},s_{i}^{B}$ times the ith A and B cycles respectively, is
\be
T_{wall}=2\pi\left|\sum_{i}(r_{A}^{i}{\cal G}_{i}+s_{i}^{B}z^{i})\right|e^{\phi/2}e^{-6u}\label{Twall}
\ee
Since ${\cal V}\sim e^{6u}$, the tension of the wall goes like $T_{wall}\sim 1/{\cal V}$. \\ 
 If such a brane is nucleated the superpotential will change by (assuming
for simplicity that the change in the complex structure moduli can be
ignored),
\be
\Delta W=\sum_{i}(r_{A}^{i}{\cal G}_{i}+s_{i}^{B}z^{i}).\label{DeltaW1}
\ee
Similarly the nucleation of an NS brane - whose action will be given
by the same formula as above but with the factor $e^{\phi/2}$ replaced
by $e^{-\phi/2}$ will cause a shift in the superpotential ,
\be
\Delta W=\sum_{i}iS(r_{A}^{i}{\cal G}_{i}+s_{i}^{B}z^{i}).\label{DeltaW2}
\ee

Such transitions however have to be accompanied by changes in the $D3$ brane charge so as to be consistent with
the tadpole cancellation condition 
\be
N_{D3}+\frac{1}{2\kappa_{10}^2T_3}\int_{\mathcal{M}_6}H_3\wedge F_3= \frac{\chi}{24},
\ee
where $N_{D3}$ is the net $D3$ charge 
and $\chi$ is the Euler number of the 4-fold of the associated F-theory.

\subsection{Review of the Large Volume Scenario}

Here we will briefly review the relevant aspects of the large volume scenario. We follow the notation and discussion of \cite{lvs}. We also set $M_{P}\equiv(8\pi G_{N})^{-1/2}=2.4\times10^{18}GeV=1$.

The superpotential, Kaehler potential and gauge kinetic function for
the theory under discussion are,

\begin{eqnarray}
W & = & W_{mod}(\Phi)+\mu(\Phi)H_{1}H_{2}+\frac{1}{6}Y_{\alpha\beta\gamma}(\Phi)C^{\alpha}C^{\beta}C^{\gamma}+\ldots,\label{eq:W}\\
K & = & K_{mod}(\Phi,\bar{\Phi})+\tilde{K}_{\alpha\bar{\beta}}(\Phi,\bar{\Phi})C^{\alpha}C^{\bar{\beta}}+[Z(\Phi,\bar{\Phi})H_{1}H_{2}+h.c.]+\ldots\label{eq:K}\\
f_{a} & = & f_{a}(\Phi).\label{eq:f_a}
\end{eqnarray}
Here $\Phi=\{\Phi^{A}\}$ and $C^{\alpha}$ are chiral superfields
(including the two Higgs doublets $H_{1,2}$) that correspond to the
moduli and MSSM/GUT fields respectively. Also
\begin{eqnarray}
K_{mod} & = & -2\ln\left({\cal V}+\frac{\xi}{2}\left(\frac{(S+\bar{S})}{2}\right)^{3/2}\right)-\ln\left(i\int\Omega\wedge\bar{\Omega}(z,\bar{z})\right)-\ln(S+\bar{S}),\label{eq:hatK}\\
W_{mod} & = & \int G_{3}\wedge\Omega+\sum_{i}A_{i}e^{-a_{r}T^{r}}.\label{eq:hatW}
\end{eqnarray}
Here ${\cal V}$ is the volume (in Einstein frame) of the internal
manifold and the $\xi$= -($\chi\zeta(3)/2(2\pi)^{3}$) term is a
correction term that is higher order in the $\alpha'$ expansion.
For typical Calabi-Yau manifolds $\xi\sim O(1)$. $S$ is the axio-dilaton,
$z=\{z^{i}\}$ represents the set of ($i=1,\ldots,h_{21}$) complex
structure moduli and $T^{r}$ ($r=1,\ldots,h_{11}$) are the (complexified)
K\"ahler moduli. The second term in (\ref{eq:hatW}) is a sum of non-perturbative terms with the parameters fixed by the condensing gauge groups (or string instantons). For simplicity below we will just consider one such term.The  Calabi-Yau manifolds that we consider
are of the `Swiss cheese' type. In the simplest such manifold consistent
with our requirements the volume may be written as%

\begin{equation}
{\cal V}=k_b\tau_{b}^{3/2}-k_s\tau_{s}^{3/2}.\label{eq:Swisscheese}
\end{equation}
In the above the tau's are K\"ahler moduli which control the volume
of the four cycles with $\tau_{b}$ effectively determining the overall
size of the CY. The $k$'s are intersection numbers of two-cycles. While in explicit calculations in the the rest of
the paper, we will use \eqref{eq:Swisscheese} for the sake of simplicity,
it should be clear from the discussion that the results would hold
even in a more general CY manifold which would allow a LVS compactification.

The potential for the moduli is (assuming that the minimum would be
at large ${\cal {\cal {\cal V}}}$ and expanding in inverse powers of it) 
\begin{eqnarray}
V & = & V_{F}+V_{D}.\label{eq:pot1}\\
V_{F} & = & \frac{4}{3}g(a|A|)^{2}\frac{\sqrt{\tau_{s}}e^{-2a\tau_{s}}}{{\cal V}}-2ga|AW_{0}|\frac{\tau_{s}e^{-a\tau_{s}}}{{\cal V}^{2}}+\frac{3}{8}\frac{\xi|W_{0}|^{2}}{g^{1/2}{\cal V}^{3}}+\ldots,\label{eq:VF}\\
V_{D} & = & \frac{f}{2}D^{2},\, D=f^{-1}k^{i}K_{i}.\label{eq:VD}
\end{eqnarray}
 Note that extremizing with respect
to $\tau_{s}$ gives us an exponentially large volume and the three
displayed terms in $V_{F}$ are all of order ${\cal V}^{-3}$. This
would mean that at the classical (negative) minimum found in
\cite{lvs}, the contribution to the F-term potential
from the dilaton and complex-structure moduli%
\footnote{At this point we ignore uplifting issues.%
} are zero at leading order in the volume expansion, since they are
$O(1/{\cal V}^{2})$. Also $V_{D}=0$ to this order in the large volume expansion, since it is positive definite
and of order $1/{\cal V}^{2}$. 
The resulting minimum is AdS (with broken SUSY) and in the next section we will first discuss transitions amongst such minima. Then we discuss the possible uplift to dS minima and their transitions. Without  loss of generality, we will set the phases of A and $W_0$ equal to zero. From now on for  convenience we will use $A$ for $|A|$ and $W_0$ for $|W_0|$ .

\section{Transitions between LVS minima}

\subsection{Case without uplift}
In this case  the LVS  potential (upto higher order terms in the volume expansion) is given by
\be
V=\frac{4}{3}g(aA)^2\frac{\sqrt{\tau_s}e^{-2a\tau_s}}{\cal{V}}-2gaAW_o\frac{\tau_se^{-a\tau_s}}{{\cal{V}}^2}+\frac{3}{8}\frac{\xi W_0^2}{\sqrt{g}{\cal{V}}^3}.
\ee
Here $\cal{V}$ is the volume of CY and $a=\frac{2\pi}{N}$, for  $SU(N)$.\\
Minimising with respect to   $\tau_s$ and $\cal{V}$, we get respectively the following equations
\ben
&&e^{-a\tau_s}=3\frac{W_0\sqrt{\tau_s}}{aA\cal{V}}\frac{a\tau_s-1}{4a\tau_s-1}=\frac{W_0}{aA\cal{V}}f(\tau_s)\ ,\label{taumin}\\
&&\frac{4}{3}\sqrt{\tau_s}f^2(\tau_s)-4\tau_s f(\tau_s)+\frac{9}{8}\frac{\xi}{\sqrt{g^3}}=0\ .\label{volmin}
\een
Here
\be
f(\tau_s)=3\frac{a\tau_s-1}{4a\tau_s-1}\sqrt{\tau_s}\ .
\ee
The value of the potential at the minima is
\be
V_0=2\frac{gW_0^2}{{\cal{V}}_0^3}\left[\frac{4}{9}\sqrt{\tau_s}f^2(\tau_s)-\frac{1}{3}\tau_s f(\tau_s)\right]=\frac{\Psi(g)}{W_0}\ ,\label{V01}
\ee
where in the second step we have again used (\ref{taumin}) and (\ref{volmin}). 

Note that $W_{0}$ depends both explicitly and implicitly on the fluxes.
The nucleation of a five brane will cause a change given by (\ref{DeltaW1})
or (\ref{DeltaW2}). While this formula does not reflect the change
due to the implicit dependence on the (stabilized values of) the complex
structure moduli and the string coupling, these should be secondary
effects. Below we will thus consider the nucleation of branes which
will effectively change the flux superpotential at the minimum; $W_{0}\rightarrow W_{0}+\Delta W_{0}$
where the change is expected to be $O(1)$.

Thus  from \eqref{V01} we have through brane nucleation,
\begin{equation}
\Delta V_{0}=-V_{0}\frac{\Delta W_{0}}{W_{0}+\Delta W_{0}}. \label{eq:DeltaV}
\end{equation}
Now from the Brown-Teitelboim analysis we have an upper bound
on the tension of the brane,
\begin{equation}
T_{ub}=\frac{|\Delta V_{0}|}{\sqrt{3V_{0}}}=\sqrt{\frac{V_{0}}{3}}\frac{\Delta W_{0}}{W_{0}+\Delta W_{0}}=\sqrt{\frac{g}{3}}\frac{|\tilde{f}(\tau_{s}(g))|}{{\cal V}_{0}^{3/2}}\frac{|W_{0}\Delta W_{0}|}{|W_{0}+\Delta W_{0}|}\ .\label{eq:Tub}
\end{equation}
Here $\tilde{f}\sim {O}(1)$ is the square root of the expression in square brackets
in (\ref{V01}). 

From  (\ref{Twall}) the tension of the brane can be written as:
\be
T=\frac{2\pi\sqrt{g}\Delta W_{0}}{{\cal V}_{0}}.
\ee
Therefore $T_{ub}> T$ implies:

\begin{equation}
\sqrt{{\cal V}_{0}}<\frac{|\tilde{f}|}{\sqrt{3}}\left|{\frac{W_{0}}{W_{0}+\Delta W_{0}}}\right|.\label{eq:volbound}
\end{equation}

Since the right hand side  is an $O(1)$ quantity this inequality is not satisfied in the LVS.  Another way to read this is that this inequality would imply  ${\cal V}_0\lesssim 1$ in string units - clearly vitiating the entire effective field theory analysis. Thus any brane nucleation take us outside our framework.
We then conclude that as long as the EFT treatment is valid AdS vacua are stable despite being non-supersymmetric. This is the main result of this section.

Next we will consider uplift of AdS minima to dS minima by adding a suitable uplift term to the potential. We will classify these uplift terms as class I and class II. Uplift terms of class I depend explicitly on $W_0$ whereas the uplift terms of class II do not explicitly depend on $W_0$.

\subsection{Class I Uplift Term}

We first consider a general uplift of the form
\be
V_{uplift}=\frac{g{W_{0}^2 d}}{{\cal{V}}^\alpha}\ .
\ee
Here $d$ is positive and independent of $\tau_s$ and $1< \alpha < 3$ in order to be able to 'uplift' the minimum of the potential to de Sitter space. 
For instance \cite{uplift1} combined D-terms and matter F-terms can induce a term of the form:
\be
V_{{\rm uplift}}=\frac{gpW_{0}^{2}}{{\cal V}^{8/3}},\label{eq:uplift3}
\ee
where $p$ is an $O(1)$ number related to the $U(1)$ charge of a
matter field living on the D3 brane. \\
We have similar uplift term from the combination of D-terms, F-terms and string loop effect,
\be
V_{{\rm uplift}}=\frac{g^{2/5}\hat{\mu}W_{0}^{2}}{{\cal V}^{14/5}}\label{eq:uplift2}
\ee
Here $\hat{\mu}$ is a complex structure dependent number which is
generated by string loop effects.\\
For most of the discussion below we will treat $\alpha$ and $d$ arbitrary. Thus the minimisation equation with respect to $\tau_s$ remains unchanged, but minimisation with respect to $\cal V$ gives
\be
-\frac{8}{3}f^2(\tau_s)\sqrt{\tau_s}+8f(\tau_s)\tau_s-\frac{9}{4}\frac{\xi}{\sqrt{g^3}}-2\alpha\frac{d}{{\cal V}^{\alpha-3}}=0\ .
\label{dV}\ee 
The potential at the minimum is 
\be
V_0=2\frac{g W_0^2}{{\cal{V}}_0^3}\Phi(\hat W_0,g)\ ,\label{Vuplifted}
\ee
with $\Phi$  given by
\be
\Phi=\frac{2}{3}(1-\alpha^{-1})f^{2}(\tau_{s})\sqrt{\tau_{s}}-(1-2\alpha^{-1})f(\tau_{s})\tau_{s}+\frac{3}{16}(1-3\alpha^{-1})\frac{\xi}{\sqrt{g^{3}}}\label{Phi}\ .
\ee
One can see that the value of $\Phi$ is bounded from below and is given as
\be
\Phi>-6\tau_{s}^{3/2}\frac{(a\tau_{s}-1)^2}{(4a\tau_{s}-1)^{2}}.\label{Phibound}
\ee


\begin{figure}[h]
\centering
\includegraphics[width=10cm]{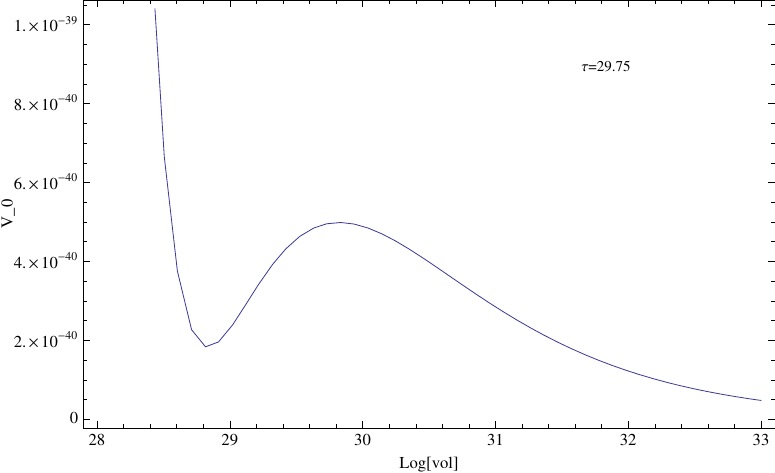}
\caption{For the uplift with $\alpha=\frac{14}{5}$, we find the de Sitter vacuum with the set of parameter $g=0.1,W_0=0.1,a=1,A=1,\xi=0.1,\hat\mu=0.1$. The minima occurs at $(\tau_s=29.75,\ln{\mathcal V}=28.81)$ and the potential at the minima is $V_0=1.83\times 10^{-40}$.}
\end{figure}

We would like to understand the decay in the landscape of de Sitter vacua. We assume for simplicity that the  the string coupling does not change under change of fluxes. As we have argued earlier there are many flux changes that keep it fixed and in any cases  the string coupling $g$ cannot change by more than a number $\lesssim O(1)$ without violating the effective field theory criteria. Similarly a brane nucleation that changes the $z$'s will have an effect on the potential minimum only to the extent that it changes $W_o$. So effectively we only need to  investigate the changes of the potential due to changes in the superpotential.

Let us start with a de Sitter vacuum. Let $(\tau_{s0},{\mathcal V_0})$ be a de Sitter minimum and the corresponding magnitude of the flux super potential be $W_0$. If we change the flux, so that $W_0\rightarrow W_0+\Delta W_0$, the location of the minimum and the value of potential at the minimum will also change. Let the new location of the minimum be $(\tau_{s0}+\Delta\tau,{\mathcal V_0}+\Delta{\mathcal V})$.\\
From (\ref{taumin}) the change of flux gives the following  relation:
\be\label{minims1}
-a\Delta\tau=\ln\left |\frac{W_0+\Delta W_0}{W_0}\right |+\frac{\Delta f}{f}-\frac{\Delta{\mathcal V}}{{\mathcal V_0}}\ .
\ee
The change in $f(\tau_s)$ can be expressed in terms of $\Delta\tau$ as
\be\label{changf2}
\frac{\Delta f}{f}=\frac{\Delta \tau}{2\tau_s(a\tau_s-1)(4a\tau_s-1)}(a\tau_s+4a^2\tau_s^2+1)\ .
\ee
Using (\ref{dV}), we get the following relation among variations
\be\label{minims3}
\alpha d (3-\alpha)\frac{\Delta{\mathcal V}}{{\mathcal V_0}^{\alpha-2}}=\frac{18a\sqrt{\tau_{s0}^3}\Delta\tau}{(4a\tau_{s0}-1)^3}J(\tau_{s0})\ .
\ee
Here $J(\tau_s)$ is
\be
J(\tau_s)=\left (12a^2\tau_s^2-11a\tau_s+5\right )\ .
\ee
From (\ref{minims3}), we see that the coefficient of $\Delta{\mathcal V}$ is positive if $\alpha<3$. Since $J$ is positive, it follows that the variations $\Delta{\mathcal V}$ and $\Delta\tau$ have the same sign. Thus if the change of flux increases $\tau_{s0}$, then the volume also increases and vice versa.\\
We substitute (\ref{changf2}) and (\ref{minims3}) in (\ref{minims1}) to get the following relation
\be\label{compr2}
\ln\left |\frac{W_0+\Delta W_0}{W_0}\right |=-\frac{\Delta\tau{\mathcal V_0}^{\alpha-3}}{2\tau_{s0}\alpha d(4a\tau_{s0}-1)(a\tau_{s0}-1)}H_\alpha(\tau_{s0})\ .
\ee
Here $H_\alpha(\tau_s)$ is given by
\ben\label{defnH}
H_\alpha(\tau_s)&=&\left(4\tau_s f-\frac{4}{3}f^2\sqrt{\tau_s}-\frac{9}{8}\frac{\xi}{\sqrt{g^3}}\right)\left(3a\tau_s+1-6a^2\tau_s^2+8a^3\tau_s^3\right)\nn\\&&-36\frac{a\tau_s^2\sqrt{\tau_s}(a\tau_s-1)}{(4a\tau_s-1)^2(3-\alpha)}(12a^2\tau_s^2-11a\tau_s+5)\ .
\een
For $\alpha<3$, in the large volume and $a\tau_s>>1$ limit, the expression for $(3-\alpha)H_\alpha$ can be approximated as
\be
(3-\alpha)H_\alpha\sim \frac{9\sqrt{\tau_s^3}}{4}(a\tau_s)\left[8(3-\alpha)a^2\tau_s^2-(30-6\alpha)a\tau_s+20-3\alpha\right]\ .
\ee


\begin{figure}[h]
\centering
\includegraphics[width=10cm]{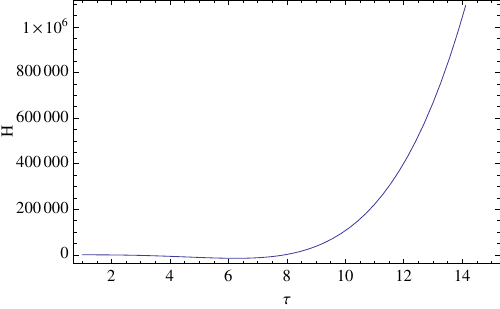}
\caption{H for $g=0.1,\alpha=\frac{14}{5}$ and $a=1$}
\label{fig2}
\end{figure}

Thus we see that for $a\tau_s>>1$,  $(3-\alpha)H_\alpha$ is a positive and monotonically increasing function. To see its behaviour for lower value of $\tau_s$, we have ploted $H_\alpha$ against $\tau_s$ in figure \ref{fig2}. 
Thus we see from (\ref{compr2}) that since $H_\alpha$ is positive for large $\tau_s$, with the decrease of $W_0$, $\tau_s$ and hence the volume ${\mathcal V}$ increases.\\ 
We can also check how the minimum of the potential changes as we change the flux. To do this we write the potential at the minimum as
\be\label{V}
V_0=\frac{2g}{W_0}(aA)^3\left(\frac{e^{-a\tau_{s0}}}{f(\tau_{s0})}\right)^3\Phi\ .
\ee
Here we have used (\ref{taumin}) to replace ${\mathcal V}$ interms of $\tau_s$ and $W_0$.\\
Thus the change in minima due to change in flux can be given as
\be\label{changeV}
\ln\left(\frac{V_0+\Delta V_0}{V_0}\right)=-\ln\left |\frac{W_0+\Delta W_0}{W_0}\right |-3a\Delta\tau-3\frac{\Delta f}{f}+\frac{\Delta\Phi}{\Phi}\ .
\ee
Here $\Phi$ is given in (\ref{Phi}). Using (\ref{changf2}), we can calculate the variation in $\Phi$,
\be\label{changePhi}
\Delta\Phi=\frac{f(\tau_s)\Delta\tau}{2\tau_s(a\tau_s-1)(4a\tau_s-1)}L(\tau_s,\alpha)\ ,
\ee
where
\be
L(\tau_s,\alpha)=2\left(1-\frac{1}{\alpha}\right)\sqrt{\tau_s}f(\tau_s)(4a^2\tau_s^2-a\tau_s+1)-3\tau_s\left(1-\frac{2}{\alpha}\right)\left[(4a^2\tau_s^2+1)-3a\tau_s\right]\ .
\ee
Substituting (\ref{changf2}),(\ref{compr2}) and (\ref{changePhi}) in (\ref{changeV}), we get following relation
\be\label{changeV2}
\ln\left(\frac{V_0+\Delta V_0}{V_0}\right)=\frac{\Delta\tau_s{\mathcal V}^{\alpha-3}}{2\alpha d\tau_s(a\tau_s-1)(4a\tau_s-1)}Q(\tau_{s0},\alpha)\ ,
\ee
with
\be
Q(\tau_s,\alpha)=-F(\tau_s)+\frac{f(\tau_s)L(\tau_s,\alpha)}{\Phi}B(\tau_s)\,
\ee
and  $B(\tau_s)$ and $F(\tau_s)$ defined as
\be\label{B}
B(\tau_s)=4\tau_s f(\tau_s)-\frac{4}{3}f^2(\tau_s)\sqrt{\tau_s}-\frac{9}{8}\frac{\xi}{\sqrt{g^3}}\ ,
\ee
\be\label{F}
F(\tau_s)=2B(\tau_s)\left(3a\tau_s+1-6a^2\tau_s^2+8a^3\tau_s^3\right)+36a\tau_s^2\sqrt{\tau_s}\frac{(a\tau_s-1)J(\tau_s)}{(4a\tau_s-1)^2(3-\alpha)}\ .
\ee
In order to see the behaviour of $Q$ we will also plot $Q$ as a function of $\tau_s$. \\


\begin{figure}[h]
\label{Q14}
\centering
\includegraphics[width=10cm]{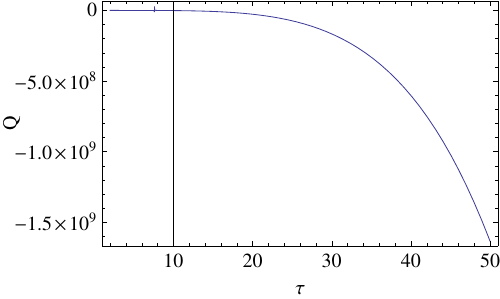}
\caption{Q for $g=0.1,\alpha=\frac{14}{5}$ and $a=1$.}
\end{figure}

Thus we see that for large $\tau_s$, Q is negative. This together with (\ref{changeV2}) and (\ref{minims3}) imply that as $\tau_s$ and $\mathcal V$ increase, the CC decreases. \\
In summary we conclude that a decrease in the magnitude of $W_0$ produces vacua with smaller vacuum energy and larger volume and $\tau_s$.

\subsection{Class II Uplift Term}

In this section we will look at the uplift of the form
\be
V_{uplift}=\frac{K}{{\mathcal V}^\alpha},\quad 1<\alpha<3\ .
\ee
Here $K$ is independent of flux. For example the uplift generated by non perturbative effect has the form \cite{uplift2}
\begin{equation}
V_{{\rm uplift}}=gh^{2}\frac{e^{-2b/g}}{{\cal V}}\label{eq:uplift1}\ .
\end{equation}
Here $b$ and $h$ are independent of $\tau_s$.\\
Thus apart from (\ref{taumin}), we get the following minimisation equation
\be
\alpha\frac{K}{gW_0^2{\mathcal V}^{\alpha-3}}=4\tau_s f(\tau_s)-\frac{4}{3}f(\tau_s)^2\sqrt{\tau_s}-\frac{9}{8}\frac{\xi}{\sqrt{g^3}}\ .
\ee
Using (\ref{taumin}), we can eliminate the volume dependence in the above equation and write it in the form
\be\label{class2uplift}
\frac{\alpha K}{gW_0^{\alpha-1}}\left(\frac{fe^{a\tau}}{aA}\right)^{3-\alpha}=4\tau f-\frac{4}{3}f^2\sqrt{\tau}-\frac{9}{8}\frac{\xi}{\sqrt{g^3}}\ .
\ee
The potential at the minimum again given by (\ref{Vuplifted}) with $\Phi$ given by (\ref{Phi}).\\
In particular in the case of uplift with $\alpha=1$, the above minimisation equation simplifies to
\be\label{minims4}
\frac{Ke^{2a\tau_s} f(\tau_s)^2}{(aA)^2}=4\tau_s f(\tau_s)-\frac{4}{3}f(\tau_s)^2\sqrt{\tau_s}-\frac{9}{8}\frac{\xi}{\sqrt{g^3}}\ .
\ee
Thus in this case if we change the flux keeping $g$ fixed, then $\tau_s$ is fixed for all such changes of flux. Since the sign of the potential at the minima depends on $\Phi(\tau_s,g)$,  if we start with de Sitter, it will  remain de Sitter as $\tau_s$ does not change with flux.\\ 
Also the potential at the minima is given by
\be
V_0=\frac{2g}{W_0}\left(\frac{e^{-a\tau_{s0}}aA}{f(\tau_{s0})}\right)^3\Phi(\tau_{s0},g)\ .
\ee
Thus we see that in the case for $\alpha=1$, the changes of flux which increases $W_0$, will reduce the CC. \\
Now let us consider cases for general $\alpha$. Taking the variation of the relation (\ref{class2uplift}) \footnote {Here we again assume that $g$ does not change so much. However in order to see the effect of the variation $g$ on $\tau$, we can use non-perturbative uplift where $\tau$ dependence on flux through $g$. In this case we get $\frac{\Delta\tau H_1}{\tau(a\tau-1)(4a\tau-1)}=\frac{\Delta g}{g^2}[\frac{27\xi}{16\sqrt{g}}-2b B]$ which in large $a\tau$ limit reduces to $\Delta\tau=-\frac{b}{a}\frac{\Delta g}{g^2}$.  }
\be
(\alpha-1)\ln\left |\frac{W_0+\Delta W_0}{W_0}\right |=(3-\alpha)\frac{\Delta f}{f}+a(3-\alpha)\Delta\tau-\frac{\Delta B}{B}\ .
\ee
$B$ is positive because of the relation (\ref{minims4}). Using (\ref{changf2}), we get the following relation
\be\label{compr3}
(\alpha-1)\ln\left |\frac{W_0+\Delta W_0}{W_0}\right |=\frac{(3-\alpha)H_\alpha}{2\tau_s(a\tau_s-1)(4a\tau_s-1)}\left(\frac{gW_0^2{\mathcal V}^{\alpha-3}}{\alpha K}\right)\Delta\tau \ .
\ee
Here $(3-\alpha)H_\alpha$ is given in (\ref{defnH}) and is positive for $\alpha<3$ and large $a\tau_s$ . Thus for $1<\alpha<3$, an increase in $W_0$ increases $\tau_s$. 
As a special case we see that for $\alpha=1$, $\Delta\tau=0$.
Comparing (\ref{compr2}) and (\ref{compr3}), we see that apart from some positive factors, the major difference is the sign.  Thus in the case of uplift of class II, we expect the opposite behaviour as a function of $W_0$ for $\tau_s,{\mathcal V}$ and $V_0$.

We also have relation

\be
(3-\alpha)(\alpha-1)\frac{\Delta{\mathcal V}}{{\mathcal V}}=\frac{1}{B(4a\tau_s-1)}\left[\frac{18a(\alpha-1)\sqrt{\tau_s^3}J}{(4a\tau_s-1)^2}+\frac{(3-\alpha)H_\alpha}{\tau_s(a\tau_s-1)}\right]\Delta\tau
\ee
For $1<\alpha<3$, the RHS is positive. Hence ${\mathcal V}$ changes in the same manner as $\tau_s$ with the change of $W_0$.\\
We also calculate the change in minimum of the potential

\be
(\alpha-1)\ln\left |1+\frac{\Delta V}{V_0}\right |=\frac{gW_0^2{\mathcal V}^{(\alpha-3)}\Delta\tau_s}{2\alpha\tau_s K(a\tau_s-1)(4a\tau_s-1)}[(\alpha-1)Q-2H_\alpha].
\ee
Now for $\alpha>1$ and large $a\tau_s$,  $[(\alpha-1)Q-2H_\alpha]$ is negative. This implies that $V_0$ decreases as $\tau_s$ and $\mathcal V$ increase.\\
In summary we conclude that in this case, in contrast to the previous case, an increase in the value of $W_0$ implies smaller vacuum energy with larger volume and $\tau_s$. 

\section{BT processes and decay rates in LVS}

In LVS we have $\Lambda\sim1/{\cal V}^{3}\,{\rm and}\, T\sim1/{\cal V}.$
Thus $\Lambda/T^{2}\sim1/{\cal V}$ is a small expansion parameter
for ${\cal V}\gg1$. We also note that the only allowed values of
$\sigma_{o/i}$ (see (\ref{signsigma})) are $\sigma_{0}=-1,\sigma_{i}=+1$. So
for $B$ we have the expression given in (\ref{Bplusminus}). Expanding this we
get,
\begin{equation}
B=\frac{24\pi^{2}}{\Lambda_{o}}+2\pi^{2}\bar{\rho}^{3}T-12\pi^{2}\bar{\rho}^{2}+\frac{\pi^{2}}{2}(\Lambda_{i}+\Lambda_{o})\bar{\rho}^{4}+O(1).\label{eq:Bexpn}
\end{equation}
Note that the first term is $O({\cal V}^{3})$ the second and third
are both of $O({\cal V}^{2})$ (since $\bar{\rho}\sim1/T\sim{\cal V}$
- see below) while the fourth term is $O({\cal V})$. Similarly expanding
the bubble radius (\ref{barrho}) we get 
\begin{eqnarray}
\bar{\rho} & = & \frac{4}{T}\left\{1-\frac{4}{3T^{2}}(\Lambda_{i}+\Lambda_{0})+O(\frac{1}{{\cal V}^{2}})\right\}.
\end{eqnarray}
Note that $\bar{\rho}T=4+O(1/{\cal V})$. Using this expansion in
\eqref{eq:Bexpn} we have,

\begin{eqnarray}
B & = & \frac{24\pi^{2}}{\Lambda_{0}}-\frac{64\pi^{2}}{T^{2}}+\frac{128\pi^{2}}{T^{4}}(\Lambda_{o}+\Lambda_{i})+O(1).\label{eq:Bapprox}
\end{eqnarray}
First note that the decay probability $P\sim e^{-B}$ is suppressed
as $\sim e^{-{\cal V}^{3}}$to leading order. Nevertheless the decay
time scale is parametrically smaller than the Poincare recurrence
time $t_{r}=e^{24\pi^{2}/\Lambda_{r}}$;
\[
t_{{\rm decay}}\sim\frac{1}{P_{{\rm decay}}}\sim e^{-\frac{64\pi^{2}}{T^{2}}}t_{r}\ll t_{r}.
\]
Let us now compute the ratio of the decay proabilities to two different
vacua with CC's $\Lambda_{i}^{(1)},\Lambda_{i}^{(2)}$. We find (with
$P_{r}\equiv\exp\{-B(\Lambda_{o}\rightarrow\Lambda_{i}^{(r)})\}$)
\begin{equation}
\frac{P_{1}}{P_{2}}=\exp\left[-\frac{128\pi^{2}}{T^{4}}(\Lambda_{i}^{(1)}-\Lambda_{i}^{(2)})\right].\label{eq:P1P2}
\end{equation}
This formula implies in particular that up trasitions are suppressed
compared to down transitions, for taking $\Lambda_{i}^{(1)}>\Lambda_{o}>\Lambda_{i}^{(2)}$
we have $P_{1}/P_{2}\sim e^{-{\cal V}}$. Similarly the decays from
dS to dS (with a lower CC ) is suppressed compared to decays from
dS to AdS since in that case $ $ (with $\Lambda_{i}^{(1)}>0$ and
$ $$\Lambda_{i}^{(2)}=-|\Lambda_{i}^{(2)}|$ so that the exponent
is again negative and $P_{1}/P_{2}\sim e^{-{\cal V}}$. 

Let us now estimate the decay to decompactification by tunneling through
the barrier in the uplifted potential. Note that this does not involve
any change in flux - it is simply a transition in the same point in
the flux landscape and is thus described by the CDL analysis. In this
case also the effective $B\sim{\cal V}^{3}$ and gives a similarly suppressed 
rate as the BT process for a flux changing decay.
This fits with the general statements of \cite{Westphal}.\\
However in this case CDL tunneling for decompactification is dominant compared to decay via BT. One can see this as follows.\\ 
In CDL analysis the tension of the bubble wall is given as
\be
T^{CDL}_{wall}=\int_{\mathcal {V}}^\infty du\sqrt{2V}\sim \frac{1}{\sqrt{\mathcal V^{3}}}< T^{\rm 5~brane}_{wall}\ .
\ee
In the above we used $\mathcal {V}=e^{6u}$. Thus the ratio of decay probabilities to decompactification and  is given by
\be
\frac{P(CDL)}{P(dS)}\sim e^{+\mathcal O(1){\mathcal V}^3}\ .
\ee
Thus the decay to another de Sitter via BT is much suppressed compared to decay via CDL tunneling to decompactification. This agrees with previous work \cite{Larfors}.
\section{Comparison with other scenarios}

In this section we will compare our results with the results that have been reported for other related scenarios of flux compactifications namely: the type IIB KKLT scenario of moduli stabilisation, non-supersymmetric type IIA flux compactifications  and (non-stringy) flux compactifications of the simple 6D Einstein-Maxwell theory \footnote{Note that in the presence of ${\bar D} 3$ branes in a more general background, for some choice of  fluxes, one can still get metastable AdS instead of dS. In this case one might wonder whether the uplifted AdS can decay via the KPV process \cite{kpv}. In the KPV process, the ${\bar D} 3$ branes expand into a fuzzy NS5 brane which has topology of $R^4\times S^2$. For a sufficiently small number of ${\bar D}3$ branes, KPV showed that this NS5 brane settles down in a metastable minimum. However this metastable minimum decays to the SUSY vacuum via the nucleation of another NS5 brane as bubble wall. However we have already shown that in AdS space (in the LVS) there is no nucleation of a NS5 brane and hence this decay will not happen in the effective field theory description.}.

\subsection{KKLT}

As mentioned in the introduction, the KKLT scenario was the first one to be considered in the context of the landscape. Even though the flux compactification part is the same as that considered in LVS (following GKP \cite{gkp} to stabilise the dilaton and complex structure moduli from fluxes), the fixing of the K\"ahler moduli  has some distinguishing features that
makes this scenario very different from the LVS.

\begin{itemize}

\item{} In KKLT the flux superpotential has to be  tuned to be very small in order that it becomes of the same order as the hierarchically  small non-perturbative superpotential. This usually requires $W_0\sim 10^{-10}$ in string units. This is very different from LVS in which the superpotential is $\lesssim  O(1)$.
The small parameter in LVS is not $W_0$ but (effectively) the inverse of the volume, which is determined dynamically to be exponentially large in the inverse string coupling at the AdS minimum before uplifting. As seen in the previous section, the transitions naturally give both the flux superpotential $W_0$ and its variation after the transition $\Delta W_0$ to be of order 1. Tuning $W_0$ and $W_0+\Delta W_0$ makes these transitions very unlikely, while in LVS  generic values of $W_0$  \cite{uniformW}\ are used, making them comparatively less suppressed.

\item{} In KKLT the original AdS minimum is supersymmetric and therefore is automatically stable. In LVS the AdS minimum before uplift is already non-supersymmetric. Therefore we needed to study carefully the potential instability of the system. We found that  as long as the EFT is trustable the decay does not occur.

\item{} The situation after uplift is similar in both cases, as long as the uplift mechanism applies also to KKLT\footnote{Notice that since vanishing F-terms imply vanishing D-terms  the AdS minimum is supersymmetric, D-term uplift does not work in KKLT but it can work in LVS.}. In particular the CDL induced decay to the decompactification vacuum is similar in both cases. But as mentioned before the difference in magnitude of the flux superpotential makes the transition  between flux vacua very different in both cases. In particular in LVS the volume increases in the transition from dS to dS whereas in KKLT the volume remains similar.

\end{itemize}

\subsection{Non-supersymmetric Type IIA  flux compactifications}

In  \cite{nt} 
 the non-perturbative decays of non-supersymmetric AdS minima in Type IIA massive theory compactified  on an orientifold of the $T^6/{(Z_3\times Z_3)}$ orbifold were analysed.  In this model there are 3 Kahler moduli, corresponding to the 3 $T^2$'s, nine blow up modes (coming from blowing up 9 singular points), the dialton-axion, and no complex structure moduli. The moduli are fixed by turning on 4-form fluxes through 4-cycles in $T^6$ and blow up cycles. In this case the decay is mediated by the nucleation of the domain wall which carries D4 brane charge. In the thin wall approximation the domain wall can be approximated as a D4 brane which wraps two cycles in the internal space and extends along 3 non-compact direction in $AdS_4$. The tension of the domain wall goes like $\frac{1}{{\mathcal V}^{13/6}}$.

 The potential at the minima goes like $\frac{1}{{\mathcal V}^3}$ and the gravitino mass goes like $\frac{1}{\sqrt{{\mathcal V}}}$ in the large volume limit. Unlike in the IIB case the dilation $e^\phi$ in IIA case goes like $\frac{1}{\sqrt{\mathcal V}}$. In their analysis \cite{nt}  found that the vacua, which are related to susy vacua by reversing the sign of all fluxes, are stable  within the context of the effective field theory analysis. This is similar to our finding for IIB  LVS vacua before uplifting. However other vacua, which are perturbatively stable, are unstable and decay via nucleation of D4 brane. The tadpole cancellation condition does not put any constraint on the 4-form charges and hence they do not play any role in their analysis. However  in our case the tadpole cancellation condition involves relation between net D3 brane charge and 3-form charge and the nucleation of domain wall changes the 3-form flux.
Finally we remark that uplifted vacua in the type IIA case have not been discussed in \cite{nt} or elsewhere as far as we know.

\subsection{Flux compactifications of 6D Einstein-Maxwell system}

In references \cite{Yang, blanco, browndahlen} a detailed study of the transitions in the simple 6D Einstein Maxwell system  compactified on a 2-sphere with non-trivial magnetic fluxes was made. The advantage of this system over string models is that it is a very simple system in which the metric is known, and shares some of the properties of more complicated  string models.
The structure of the vacua is determined by three contributions to the scalar potential: the contribution from curvature of the compact two-sphere, the positive contribution from the flux of the E$\&$M field and the original positive  6D cosmological constant. Depending on the value of the fluxes the minimum is dS, Minkowski or AdS. For large fluxes the vacuum is dS, reducing the value of the quantised fluxes reduces the value of the vacuum energy until it moves to Minkowski, then AdS and finally in the absence of fluxes the potential becomes unbounded from below. This last stage is interpreted as a bubble of nothing transition by \cite{browndahlen}. Reducing the value of the fluxes not only reduces the value of the vacuum energy but also the value of the radius of the extra dimensions. However the limit towards zero flux gives very small values of the extra dimensions for which the effective field theory is not  valid.

In LVS the scalar potential also has at least three terms but of different origin, since there are neither curvature terms nor a higher dimensional cosmological constant.
The terms include the positive term coming from the nonperturbative superpotential $W_{np}$ that by itself gives a runaway to infinite volume, the negative term coming from the combined non-perturbative and flux superpotentials,  the $\alpha'$ corrections to the K\"ahler potential which provides a positive term proportional to the square of the flux superpotential $W_0\gg W_{np}$, and finally a fourth positive term which provides the dS uplift. Note that even without the uplift term the scalar potential goes to zero from above for large volume. Using the flux superpotential $W_0$ as a parameter (instead of the fluxes themselves) we can easily see that for a a relatively large $|W_0|$ the volume is large ${\cal V}\sim W_0 e^{a/g}$  while the vacuum energy goes like $|W_0|^2/{\cal V}^3$.

Reducing $|W_0|$ (keeping g fixed) the vacuum energy gets smaller as does the volume, in the AdS regime, but since $V_0\sim |W_0|^2/{\mathcal V}^3\sim 1/|W_0|$ the magnitude of the vacuum energy increases linearly with the reduction of $W_0$. However for a critical value of $|W_0|$, when it gets of the order of the non-perturbative superpotential $W_0\sim W_{np}$ the volume has no significant dependence on $W_0$ and then the vacuum energy becomes of order $V_0\sim |W_0|^2$ as in KKLT.

Therefore instead of continuing to become a  deeper AdS as in the 6D case, $V_0$ reverses direction and grows to less negative values \cite{shehu}. In the limit of zero fluxes $|W_0| \to 0$ the potential shows the runaway behaviour of the pure (positive) non-perturbative effect. Therefore, contrary to the 6D case (in which zero fluxes leads to unbounded from below potential), there is no indication of a bubble of nothing decay. Of course once the volume at the minimum reaches values of order the string scale, the effective field theory description loses meaning and a definite statement regarding the behaviour of the system close to vanishing size of the extra dimensions, as happens in the bubble of nothing, is beyond our theoretical framework.


\begin{figure}[h]
\label{lvsvs6d}
\centering
\includegraphics[width=10cm]{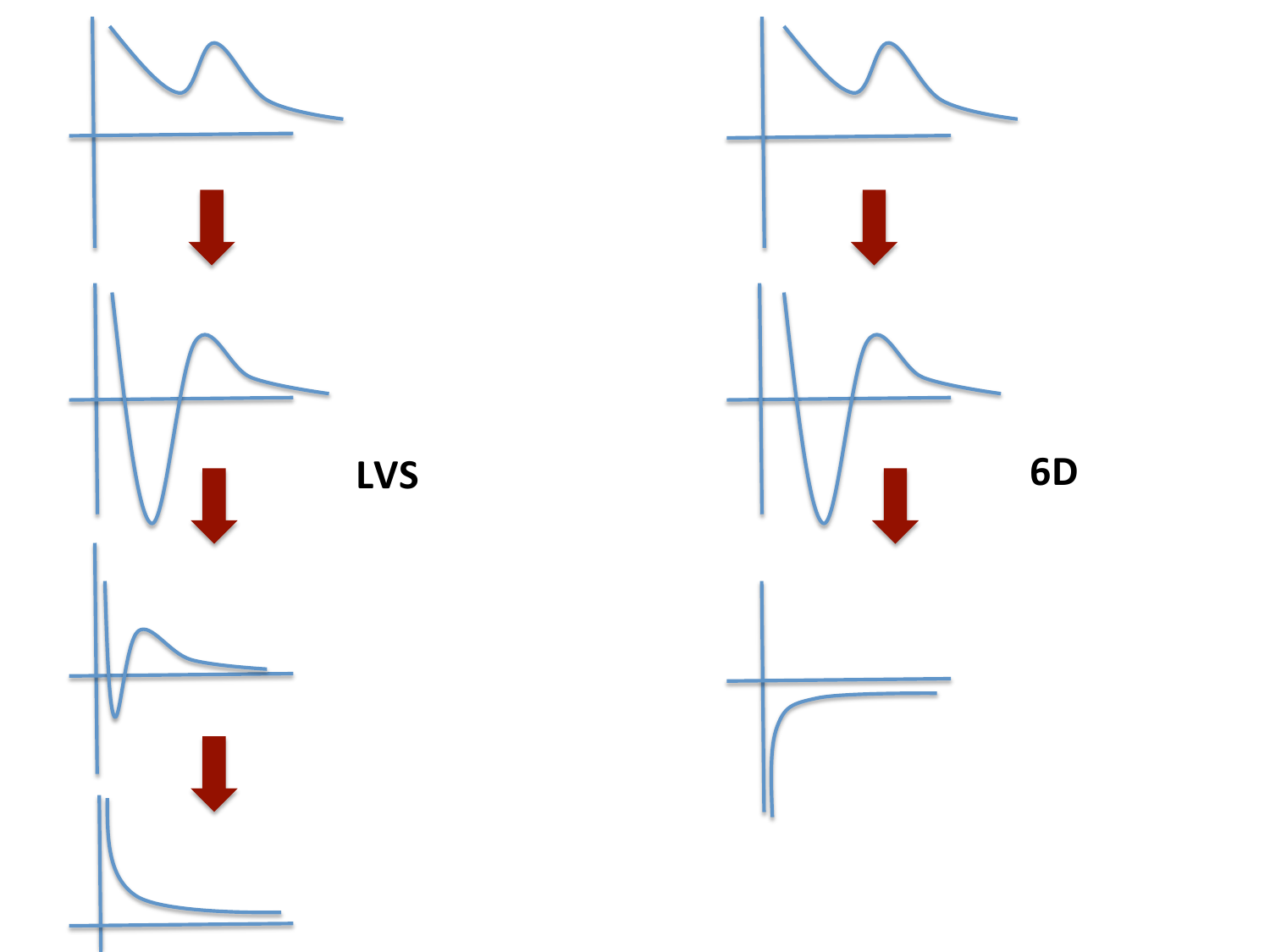}
\caption{Change of the vacuum energy and value of the volume (horizontal axis) at the minimum as fluxes change for the LVS (left) and the 6D Einstein-Maxwell system (right). Notice that in both cases the vacuum energy and volume are initially reduced but even though in the 6D case they continue decreasing until reaching a potential unbounded from below (bubble of nothing) in the absence of fluxes, in the LVS case the process is reversed and the vacuum energy gets bigger after a critical value leading to the runaway behaviour in the absence of fluxes. Warning: the arrows do not mean transitions, only changes when the superpotential is reduced.}
\end{figure}

\section{Conclusions}

In this article we have discussed the non-perturbative stability of the large volume scenario.  This is in general relevant for discussions related to the landscape of IIB flux compactifications since this is the scenario applicable for typical values of the flux superpotential.
We have arrived at a few general conclusions. In particular we established that the unlifted AdS vacuum is stable (within the effective field theory context) even though it corresponds to spontaneously broken supersymmetry.  This opens the possibility that these vacua may have CFT duals (since an instability could be seen as departure from conformality). It would be interesting to identify and characterise these potential duals. 

The behaviour of the dS minima is more model dependent depending crucially on the uplifting mechanism. In the more studied KKLT case the uplifting by anti D3 branes in a warped region has been the standard for studies of the landscape and its population. The LVS allows for different uplifting mechanisms and we classified them into two general classes that capture the main proposed scenarios. In all cases we found that  either increasing or decreasing  the flux superpotential gives rise to smaller vacuum energies and volumes. We computed the decay rates from dS to dS and dS to the decompactified minimum using the Brown-Teitelboim and Coleman-De Luccia formalisms.  The results can be captured in exponentials of powers of the volume. Roughly each rate is suppressed by approximately $e^{{-\cal V}^3+...}$.  The leading order in the exponential gives the inverse of the Poincar\'e recurrence time. It is then important that the next-to-leading order term in the exponential is positive to make the decay rate much faster than the Poincar\'e recurrence time. This is indeed the case. Then the relevant quantities are the relative ratios. 

We found that the ratio of probabilities from a given dS minimum towards negative cosmological constant (big crunch sinks)  $P_{bc}$ dominates over the probability of decay to another dS minimum  $P_{dS}$ by a factor of order $P_{bc}/P_{dS}\sim e^{\cal V}$. A similar behaviour has been found in other context within the landscape \cite{sinkskklt,maldacena}.  Also within dS transitions, decays towards larger volume and smaller vacuum energy are dominant,  proportional to $e^{\cal V}$. Finally decays to dS are sub-dominant compared to  the CDL transition to decompactification by a factor of order $e^{-{\cal V}^3}$.  This quantification of ratios of probabilities as functions of the volume establishes an interesting hierarchy of decays that should be useful in global studies of the dynamics of the landscape.

 We compared our results with studies on IIA compactifications as well as KKLT and 6D Maxwell-Einstein and found similarities but also important differences with all of them. In particular, contrary to the 6D case, we found no indications for decay to a bubble of nothing. However a proper stringy study of the potential of a bubble of nothing decay is beyond the scope of this article.

The fact that the generic values of the flux superpotential are of order ${\cal O}(1)$ as is the difference of the flux superpotentials between two different vacua, makes the detailed study of the LVS in any discussion of the landscape of IIB flux compactifications worth considering. 

 Implications about the age of de Sitter and the appearance of the scale of supersymmetry breaking in studies of the landscape may have to be reconsidered. The fact that the transitions change the fluxes and then also the effective number of D3 branes should have interesting phenomenological and cosmological implications. The location of the D3 branes induced in this process and their physical implications deserve further study.
We also hope this work will be useful for further formal studies of the string landscape, including the measure problem, bounds on de Sitter lifetime, etc.

\subsection*{Acknowledgements}

We thank Joe Polchinski for a question that prompted this investigation. We also thank useful discussions with Cliff Burgess, Michele Cicoli, Joseph Conlon, Anshuman Maharana, Roberto Valandro and Alexander Westphal. SdA thanks the  Abdus Salam International Centre for Theoretical Physics (ICTP)  for  the award of a visiting professorship. The  research of SdA was supported in part  by the United States Department of Energy under grant DE-FG02-91-ER-40672.


\begin{thebibliography}{99}


\bibitem{cdl}
  S.~R.~Coleman and F.~De Luccia,
  ``Gravitational Effects on and of Vacuum Decay,''
  Phys.\ Rev.\ D {\bf 21} (1980) 3305.

\bibitem{bt}
  J.~D.~Brown and C.~Teitelboim,
  ``Neutralization of the Cosmological Constant by Membrane Creation,''
  Nucl.\ Phys.\ B {\bf 297} (1988) 787.


\bibitem{witten}
E.~Witten,
  ``Instability of the Kaluza-Klein Vacuum,''
  Nucl.\ Phys.\ B {\bf 195} (1982) 481.

\bibitem{kklt}
  S.~Kachru, R.~Kallosh, A.~D.~Linde and S.~P.~Trivedi,
  ``De Sitter vacua in string theory,''
  Phys.\ Rev.\ D {\bf 68} (2003) 046005
  [hep-th/0301240].



\bibitem{lvs}
 V.~Balasubramanian, P.~Berglund, J.~P.~Conlon and F.~Quevedo,
  ``Systematics of moduli stabilization in Calabi-Yau flux compactifications,''
  JHEP {\bf 0503} 007 (2005)
  [arXiv:hep-th/0502058];
 J.~P.~Conlon, F.~Quevedo and K.~Suruliz,
  ``Large-volume flux compactifications: Moduli spectrum and D3/D7 soft supersymmetry breaking,''
  JHEP {\bf 0508} 007 (2005)
  [arXiv:hep-th/0505076].

\bibitem{Conlon} 
  J.~P.~Conlon and F.~G.~Pedro,
  ``Moduli-Induced Vacuum Destabilisation,''
  JHEP {\bf 1105}, 079 (2011)
  [arXiv:1010.2665 [hep-th]];
  D.~-i.~Hwang, F.~G.~Pedro and D.~-h.~Yeom,
  ``Moduli destabilization via gravitational collapse,''
  arXiv:1306.6687 [hep-th].

\bibitem{Hartle} 
  J.~B.~Hartle, S.~W.~Hawking and T.~Hertog,
  ``No-Boundary Measure of the Universe,''
  Phys.\ Rev.\ Lett.\  {\bf 100}, 201301 (2008)
  [arXiv:0711.4630 [hep-th]]; D.~-i.~Hwang, H.~Sahlmann and D.~-h.~Yeom,
  ``The No-boundary measure in scalar-tensor gravity,''
  Class.\ Quant.\ Grav.\  {\bf 29}, 095005 (2012)
  [arXiv:1107.4653 [gr-qc]]; D.~-i.~Hwang, B.~-H.~Lee, H.~Sahlmann and D.~-h.~Yeom,
  ``The no-boundary measure in string theory: Applications to moduli stabilization, flux compactification, and cosmic landscape,''
  Class.\ Quant.\ Grav.\  {\bf 29}, 175001 (2012)
  [arXiv:1203.0112 [gr-qc]].

 \bibitem{kpv}
  S.~Kachru, J.~Pearson and H.~L.~Verlinde,
  ``Brane / flux annihilation and the string dual of a nonsupersymmetric field theory,''
  JHEP {\bf 0206} (2002) 021
  [hep-th/0112197].
  C.~M.~Brown and O.~DeWolfe,
  ``Brane/flux annihilation transitions and nonperturbative moduli stabilization,''
  JHEP {\bf 0905} (2009) 018
  [arXiv:0901.4401 [hep-th]].
  
 \bibitem{sinkskklt}
  A.~Ceresole, G.~Dall'Agata, A.~Giryavets, R.~Kallosh and A.~D.~Linde,
  ``Domain walls, near-BPS bubbles, and probabilities in the landscape,''
  Phys.\ Rev.\ D {\bf 74} (2006) 086010
  [hep-th/0605266];
  A.~D.~Linde,
  ``Sinks in the Landscape, Boltzmann Brains, and the Cosmological Constant Problem,''
  JCAP {\bf 0701} (2007) 022
  [hep-th/0611043].
  
   \bibitem{maldacena}
  J.~Maldacena,
  ``Vacuum decay into Anti de Sitter space,''
  arXiv:1012.0274 [hep-th].
 
\bibitem{Yang} 
  I-S.~Yang,
  ``Stretched extra dimensions and bubbles of nothing in a toy model landscape,''
  Phys.\ Rev.\ D {\bf 81}, 125020 (2010)
  [arXiv:0910.1397 [hep-th]].
  
   \bibitem{blanco}
  J.~J.~Blanco-Pillado, D.~Schwartz-Perlov and A.~Vilenkin,
  ``Quantum Tunneling in Flux Compactifications,''
  JCAP {\bf 0912} (2009) 006
  [arXiv:0904.3106 [hep-th]];
  J.~J.~Blanco-Pillado and B.~Shlaer,
  ``Bubbles of Nothing in Flux Compactifications,''
  Phys.\ Rev.\ D {\bf 82} (2010) 086015
  [arXiv:1002.4408 [hep-th]];
  J.~J.~Blanco-Pillado, H.~S.~Ramadhan and B.~Shlaer,
  ``Decay of flux vacua to nothing,''
  JCAP {\bf 1010} (2010) 029
  [arXiv:1009.0753 [hep-th]].
  
  
  \bibitem{browndahlen}
  A.~R.~Brown and A.~Dahlen,
  ``Small Steps and Giant Leaps in the Landscape,''
  Phys.\ Rev.\ D {\bf 82} (2010) 083519
  [arXiv:1004.3994 [hep-th]];
  ``Giant Leaps and Minimal Branes in Multi-Dimensional Flux Landscapes,''
  Phys.\ Rev.\ D {\bf 84} (2011) 023513
  [arXiv:1010.5241 [hep-th]];
  ``Bubbles of Nothing and the Fastest Decay in the Landscape,''
  Phys.\ Rev.\ D {\bf 84} (2011) 043518
  [arXiv:1010.5240 [hep-th]];
  ``On 'nothing' as an infinitely negatively curved spacetime,''
  Phys.\ Rev.\ D {\bf 85} (2012) 104026
  [arXiv:1111.0301 [hep-th]];
  ``Populating the Whole Landscape,''
  Phys.\ Rev.\ Lett.\  {\bf 107} (2011) 171301
  [arXiv:1108.0119 [hep-th]].

  \bibitem{banks}
  T.~Banks,
  ``Heretics of the false vacuum: Gravitational effects on and of vacuum decay. 2.,''
  hep-th/0211160.
  ``The Top $10^{500}$ Reasons Not to Believe in the Landscape,''
  arXiv:1208.5715 [hep-th].
  
  \bibitem{shanta}
  S.~P.~de Alwis,
  ``Transitions Between Flux Vacua,''
  Phys.\ Rev.\ D {\bf 74} (2006) 126010
  [hep-th/0605184].
  
  
 \bibitem{polchinski}
  J.~Polchinski,
  ``The Cosmological Constant and the String Landscape,''
  hep-th/0603249.

 \bibitem{frey}
  A.~R.~Frey, M.~Lippert and B.~Williams,
  ``The Fall of stringy de Sitter,''
  Phys.\ Rev.\ D {\bf 68} (2003) 046008
  [hep-th/0305018].
  
  \bibitem{gkp}
  S.~B.~Giddings, S.~Kachru and J.~Polchinski,
  ``Hierarchies from fluxes in string compactifications,''
  Phys.\ Rev.\ D {\bf 66} (2002) 106006
  [hep-th/0105097].
  
  \bibitem{sda}
  S.~P.~de Alwis,
  ``The Scales of brane nucleation processes,''
  Phys.\ Lett.\ B {\bf 644} (2007) 77
  [hep-th/0605253].
  
  \bibitem{Kachru:2002ns}
  S.~Kachru, X.~Liu, M.~B.~Schulz and S.~P.~Trivedi,
  ``Supersymmetry changing bubbles in string theory,''
  JHEP {\bf 0305} (2003) 014
  [hep-th/0205108].
  
  \bibitem{uplift1}
M.~Cicoli, S.~Krippendorf, C.~Mayrhofer, F.~Quevedo and R.~Valandro,
  ``D-Branes at del Pezzo Singularities: Global Embedding and Moduli Stabilisation,''
  JHEP {\bf 1209} (2012) 019
  [arXiv:1206.5237 [hep-th]];
``D3/D7 Branes at Singularities: Constraints from Global Embedding and Moduli Stabilisation,''
  arXiv:1304.0022 [hep-th].

  \bibitem{uplift2}
  M.~Cicoli, A.~Maharana, F.~Quevedo and C.~P.~Burgess,
  ``De Sitter String Vacua from Dilaton-dependent Non-perturbative Effects,''
  JHEP {\bf 1206} (2012) 011
  [arXiv:1203.1750 [hep-th]].

  
  \bibitem{Westphal} 
  A.~Westphal,
  ``Lifetime of Stringy de Sitter Vacua,''
  JHEP {\bf 0801}, 012 (2008)
  [arXiv:0705.1557 [hep-th]].
  
  \bibitem{Larfors}
  M.~C.~Johnson and M.~Larfors,
  ``An Obstacle to populating the string theory landscape,''
  Phys.\ Rev.\ D {\bf 78} (2008) 123513
  doi:10.1103/PhysRevD.78.123513
  [arXiv:0809.2604 [hep-th]];
  A.~Aguirre, M.~C.~Johnson and M.~Larfors,
  ``Runaway dilatonic domain walls,''
  Phys.\ Rev.\ D {\bf 81} (2010) 043527
  doi:10.1103/PhysRevD.81.043527
  [arXiv:0911.4342 [hep-th]].
  
  \bibitem{uniformW}
  F.~Denef and M.~R.~Douglas,
  ``Distributions of flux vacua,''
  JHEP {\bf 0405} (2004) 072
  [hep-th/0404116];
  J.~J.~Blanco-Pillado, M.~Gomez-Reino and K.~Metallinos,
  ``Accidental Inflation in the Landscape,''
  JCAP {\bf 1302} (2013) 034
  [arXiv:1209.0796 [hep-th]];
  D.~Martinez-Pedrera, D.~Mehta, M.~Rummel and A.~Westphal,
  ``Finding all flux vacua in an explicit example,''
  JHEP {\bf 1306} (2013) 110
  [arXiv:1212.4530 [hep-th]].

  
  \bibitem{nt}
  P.~Narayan and S.~P.~Trivedi,
  ``On The Stability Of Non-Supersymmetric AdS Vacua,''
  JHEP {\bf 1007} (2010) 089
  [arXiv:1002.4498 [hep-th]].
  
   \bibitem{shehu}
  S.~S.~AbdusSalam, J.~P.~Conlon, F.~Quevedo and K.~Suruliz,
  ``Scanning the Landscape of Flux Compactifications: Vacuum Structure and Soft Supersymmetry Breaking,''
  JHEP {\bf 0712} (2007) 036
  [arXiv:0709.0221 [hep-th]].

  
  
    
      
   
   
   
\end{thebibliography}
\end{document}